\documentclass[a4paper,11pt]{article}
\pdfoutput=1
\usepackage{jheppub}
\usepackage{amsmath,amssymb,bm,graphicx,bbold,epsf,colordvi}
\usepackage{slashed}
\def\bea#1\eea{\begin{align}#1\end{align}} 
\usepackage{diagbox}
\usepackage{lipsum}
\usepackage{soul}
\usepackage{braket}
\usepackage{bm}
\allowdisplaybreaks 
\usepackage{color}
\usepackage[dvipsnames]{xcolor}

\usepackage{mathrsfs}
\usepackage{appendix}
\usepackage{bm}
\usepackage{lipsum}

\newcommand{\bef}{\begin{figure}[hbt]\centering}
\newcommand{\eef}{\end{figure}}

\usepackage{mathtools}
\usepackage{subfigure}
\usepackage{booktabs}
\usepackage{graphicx}
\graphicspath{ {./figure/} }
\usepackage{caption}
\usepackage{lipsum}

\newcommand{\nnu}{\nonumber\\}

\newcommand{\beq}{\begin{equation}}
\newcommand{\eeq}{\end{equation}}
\def\bea#1\eea{\begin{align}#1\end{align}}

\def \be  {\begin{equation}}
\def \ee  {\end{equation}}
\def \ba  {\begin{eqnarray}}
\def \ea  {\end{eqnarray}}

\allowdisplaybreaks

\usepackage{graphicx}

\usepackage{stackengine}

\title{Quantum Simulation of Chiral Phase Transitions}

\author[a,c]{Alexander M. Czajka}
\author[a,b,c,d]{, Zhong-Bo Kang}
\author[a,c,e]{, Henry Ma}
\author[a,b,c]{, Fanyi Zhao}

\affiliation[a]{Department of Physics and Astronomy, University of California, Los Angeles, CA 90095, USA}
\affiliation[b]{Mani L. Bhaumik Institute for Theoretical Physics, University of California, Los Angeles, CA 90095, USA}
\affiliation[c]{Center for Quantum Science and Engineering, University of California, Los Angeles, CA 90095, USA}
\affiliation[d]{Center for Frontiers in Nuclear Science, Stony Brook University, Stony Brook, NY 11794, USA}
\affiliation[e]{Department of Computer Science, University of California, Los Angeles, CA 90095, USA}

\emailAdd{aczajka74@physics.ucla.edu}
\emailAdd{zkang@ucla.edu}
\emailAdd{hetm@g.ucla.edu}
\emailAdd{fanyizhao@physics.ucla.edu}

\abstract{The Nambu-Jona-Lasinio (NJL) model has been widely studied for investigating the chiral phase structure of strongly interacting matter. The study of the thermodynamics of field theories within the framework of Lattice Field Theory is limited by the sign problem, which prevents Monte Carlo evaluation of the functional integral at a finite chemical potential. Using the quantum imaginary time evolution (QITE) algorithm, we construct a quantum simulation for the $(1+1)$ dimensional NJL model at finite temperature and finite chemical potential. We observe consistency among digital quantum simulation, exact diagonalization and analytical solution, indicating further applications of quantum computing in simulating QCD thermodynamics. }

\begin{document}

\maketitle

\section{Introduction}
The current generation of Noisy Intermediate-Scale Quantum (NISQ) technology has demonstrated that quantum computers can solve difficult problems such as simulating real-time dynamics \cite{Chiesa2019,Smith2019,Zhang2017,doi:10.1126/science.1232296,PhysRevB.101.014411,Feynman,doi:10.1126/science.273.5278.1073,DeJong:2020riy,deJong:2021wsd}, modelling relativistic behaviors and many-body systems \cite{wallraff_strong_2004,majer_coupling_2007,jordan_quantum_2012,zohar_simulating_2012,zohar_cold-atom_2013,banerjee_atomic_2013,banerjee_atomic_2012,wiese_ultracold_2013,wiese_towards_2014,jordan_quantum_2014,garcia-alvarez_fermion-fermion_2015,marcos_two-dimensional_2014,bazavov_gauge-invariant_2015,zohar_quantum_2015,mezzacapo_non-abelian_2015,dalmonte_lattice_2016,zohar_digital_2017,martinez_real-time_2016,bermudez_quantum_2017,gambetta_building_2017,krinner_spontaneous_2018,macridin_electron-phonon_2018,zache_quantum_2018,zhang_quantum_2018,klco_quantum-classical_2018,klco_digitization_2019,gustafson_quantum_2019,nuqs_collaboration_ensuremathsigma_2019,magnifico_real_2020,jordan_quantum_2019,lu_simulations_2019,klco_minimally-entangled_2020,lamm_simulation_2018,klco_su2_2020,alexandru_gluon_2019,mueller_deeply_2020,lamm_parton_2020,chakraborty_digital_2020,Bermudez:2018eyh,Ziegler:2020zkq,Ziegler:2021yua}, estimating ground states \cite{Arute:2020uxm,Ma2020,Kandala2019,PhysRevX.6.031007,Kandala2017,Peruzzo2014,PhysRevX.8.011021}, and determining finite-temperature properties of various systems \cite{Bauer,PhysRevA.61.022301,PhysRevLett.103.220502,PhysRevLett.108.080402,Temme2011,Yung754,Li:2021kcs,Zhang:2020uqo}. Though digital quantum simulations on thermal physical systems have been studied in early literature, finite-temperature physics is less understood and remains to be developed on quantum computers \cite{PRXQuantum.2.010317}. In recent years, various algorithms for imaginary time evolution on quantum computers with or without ansatz dependency have been introduced. The Quantum Imaginary Time Evolution (QITE) algorithm \cite{Motta,McArdle,Verein,PhysRevB.100.094434,Nishi,Gomes,Yeter-Aydeniz21,Yeter-Aydeniz}, an iterative technique that uses a unitary operation to simulate imaginary time evolution, has been used to calculate finite-temperature observables such as energy \cite{PRXQuantum.2.010317,Ville:2021hrl}, magnetization in the Transverse Field Ising Model (TFIM) \cite{Ville:2021hrl}, etc.

As described in Quantum ChromoDynamics (QCD), the strongly interacting matter is believed to have a complex phase structure, which is of great importance in theoretical physics~\cite{PhysRevD.77.114028,COSTA2007431,PhysRevD.77.014006,PhysRevD.91.056003,Jiang2011,PhysRevD.90.114031,Shi2014}. In the infinite quark limit, one has the deconfinement phase transition~\cite{Ratti:2021ubw} as a function of temperature as a result of the spontaneous $\mathbb{Z}_N$ symmetry breaking~\cite{Holland:2000uj}, while in the massless quark limit, one finds an order parameter for the chiral phase transition~\cite{Rajagopal:1995bc}, which has been extensively studied in~\cite{Borsanyi:2010cj,HotQCD:2018pds,Borsanyi:2020fev,HotQCD:2019xnw,Ding:2020xlj} \footnote{Further phases of strongly interacting matter are hypothesized to exist, such as the Color-Flavor Locked (CFL) phase, which occupy in the high density, low temperature region of the phase diagram.}. Because QCD is dominated by non-perturbative effects at low energies, it is difficult to study the chiral phase transition of strongly interacting matter.

To investigate the structure of the QCD chiral phase diagram\footnote{That is: the division of the temperature-chemical potential plane into regions based on the qualitative physical properties of strongly interacting matter, ``phases'', with these phases coexisting at boundaries.}, calculations are typically carried out in lattice QCD. However, the utility of this traditional approach is limited by the fermion sign problem. At nonzero baryochemical potential $\mu$, the (Euclidean) QCD action $S$ is no longer constrained to be real, so that the quantity $e^{S}$ can no longer be used as a weight for a Monte Carlo evaluation of thermal expectation values \cite{Ratti:2021ubw, Philipsen:2007rj}. The traditional approach to this problem involves simulations on imaginary chemical potentials~\cite{deForcrand:2002hgr,DElia:2002tig,Wu:2006su,DElia:2007bkz,Conradi:2007be,deForcrand:2008vr,DElia:2009pdy,Moscicki:2009id} and expansion of the expectation value $\langle\mathcal O\rangle$ of a given observable $\mathcal O$ in powers of $\mu/T$~\cite{Allton:2002zi,Allton:2005gk,Gavai:2008zr,MILC:2008reg,Kaczmarek:2011zz}, namely $\langle\mathcal O\rangle = \sum_k \mathcal{O}_k(T)(\mu/T)^k$, and evaluating the coefficients $\mathcal{O}_k(T)$ on the lattice \cite{Ratti:2021ubw, Allton:2005gk,Philipsen:2007rj}. While as pointed out in \cite{Ratti:2021ubw}, noisy results emerge in the higher order coefficients of the Taylor series from the direct simulation. In addition to the fact that the domain of validity of such an approach is limited to a (potentially small) range of baryochemical potentials, it assumes that for fixed $T$ the expectation values $\langle\mathcal O\rangle$ are analytic in $\mu$, an assumption that should fail for at least some observable in the vicinity of a finite-order phase transition. The sign problem is thus a significant obstacle to understanding the phase structure of nuclear matter. However, this is not a defect of the underlying theory (QCD), but of the attempt to simulate quantum statistics (via the functional integral) with a classical algorithm (Markov Chain Monte Carlo). By modeling the lattice system on a quantum computer, one can leverage the statistical properties \textit{of the computer itself} to eliminate the need for a Monte Carlo analysis. The fact that quantum simulations can avoid the sign problem has also been pointed out in \cite{Kharzeev:2020kgc}\footnote{Evidently, the idea of using quantum systems to simulate other quantum systems was first suggested by Feynman \cite{Feynman}.}.

To study the QCD chiral phase transition, the Nambu-Jona-Lasinio (NJL) model \cite{Nambu:1961tp,Nambu:1961fr} has been used as a convenient and practical tool due to the easy access to the dynamical mass and the mechanism of chiral symmetry breaking~\cite{PhysRevD.77.114028,COSTA2007431,Lu2015,doi:10.1142/S0217751X15501997,CUI2015172,PhysRevD.88.114019,PhysRevD.91.036006,PhysRevC.80.065805,PhysRevC.79.035807,PhysRevC.75.015805,PhysRevD.86.071502}. The NJL model, an effective model for low-energy two-flavor QCD, is also amenable to analytical calculations at finite temperature and chemical potential. A simplified version of the NJL model, the Gross-Neveu (GN) model \cite{Gross:1974jv}, is a renormalizable and asymptotically free $(1+1)$-dimensional theory comprised of $N$ fermion species interacting via four-fermion contact interaction, and it is known to exhibit chiral symmetry breaking at low energy through the generation of a nonvanishing chiral condensate $\langle\bar\psi\psi\rangle$. We will analyze the chiral condensate of the GN model at finite temperature and chemical potential \cite{Thies:2019ejd} and compare the results to quantum simulations.

In this work, we use a 4-qubit system to simulate the $(1+1)$ dimensional NJL Hamiltonian using single-qubit gates and the CNOT gate, via the Jordan-Wigner transformation \cite{Jordan1928}. In Sec.~\ref{sec:theory}, we provide analytical calculations for the finite temperature behavior of the NJL model including chemical potentials. A description of the discretization of the field theory on a lattice and the quantum algorithm we use, the QITE algorithm, is given in Sec.~\ref{sec:njlqite}. Then we present a comparison between the results obtained by analytical calculation, exact diagonalization, and quantum simulation at various chemical potentials in Sec.~\ref{sec:pheno}, where a strong consistency is found. We summarize our results in Sec.~\ref{sec:conclusion}, and discuss further directions of using quantum computing for studying finite-temperature properties of QCD.

\section{Nambu-Jona-Lasinio model in $1+1$ dimensions}
\label{sec:theory}
In this section, we provide an overview for the Nambu-Jona-Lasinio model and its phase transition at finite temperature and finite chemical potential. The Lagrangian density of the Nambu-Jona-Lasinio (NJL) model in $(1+1)$-dimensional Minkowski space is \cite{Nambu:1961tp,Nambu:1961fr}
\bea
\label{eq:Lagrangian1}
\mathcal{L}_{\rm NJL}&=\bar{\psi}(i\slashed{\partial}-m)\psi+g\left[(\bar{\psi}\psi)^2+(\bar{\psi}i\gamma_5\tau_a\psi)^2\right]\,,
\eea
where $\slashed{\partial}=\gamma_\mu\partial^\mu$, $\tau_a$ are the Pauli matrices in the isospin space, $m$ is the bare quark mass and $g$ is the dimensionless coupling constant. In $(1+1)$ dimensions, the gamma matrices $\gamma_0,\ \gamma_1$ and $\gamma_5$ are given by,
\bea
\gamma_0=\sigma_Z,\quad\gamma_1=-i\sigma_Y,\quad\gamma_5=\gamma_0\gamma_1=-\sigma_X\,,\label{eq:q5}
\eea
where $\sigma_X$, $\sigma_Y$ and $\sigma_Z$ are Pauli matrices. 
A simplified variant of the NJL model is the Gross-Neveu (GN) model \cite{Gross:1974jv} with the following Lagrangian 
\bea
\label{eq:Lagrangian2}
\mathcal{L_{\rm GN}}&=\bar{\psi}(i\slashed{\partial}-m)\psi+g(\bar{\psi}\psi)^2\,.
\eea
In the chiral limit $m = 0$, this Lagrangian is symmetric under left-action by the discrete symmetry group $\mathbb{Z}_{2, L}\times\mathbb{Z}_{2, R}\equiv\{\pm 1, \pm \gamma_5\}$ on the fields, i.e. $\psi \mapsto G\cdot \psi$ for any group element $G\in\mathbb{Z}_{2, L}\times\mathbb{Z}_{2, R}$.

In this work, we study the chiral phase transition of GN type model with non-zero chemical potential $\mu$ \cite{Thies:2019ejd},
\bea
\label{eq:Lagrangian}
\mathcal{L}&=\bar{\psi}(i\slashed{\partial}-m)\psi+g(\bar{\psi}\psi)^2+\mu\bar{\psi}\gamma_0\psi\,,
\eea
which has been applied for studying the chiral phase transition at finite chemical potential $\mu$ and temperature $T$. Throughout this paper, we refer to the  Lagrangian in Eq.~\eqref{eq:Lagrangian} as we mention ``NJL model''. 

To distinguish between different phases, we study the chiral condensate $\langle\bar\psi\psi\rangle$ in the vacuum, which is known to be an order parameter~\cite{RevModPhys.53.43,Nambu:1961tp,PhysRevD.24.450,POLYAKOV1978477,Fang:2018vkp} in the chiral limit with zero quark mass. Since the quantity $\bar\psi\psi$ transforms nontrivially under $\mathbb Z_{2,L}\times\mathbb Z_{2,R}$, a nonzero value of $\langle\bar\psi\psi\rangle$ requires spontaneous breaking of this symmetry. Thus, the transition from nonvanishing to vanishing values of the condensate signals a chiral phase transition. Thus, with non-zero quark mass, the chiral condensate $\langle\bar\psi\psi\rangle$ can be regarded as a quasi-order parameter~\cite{PhysRevD.98.094501}. 

The transition between the chirally symmetric and broken phases can be analyzed in the \textit{mean field} approximation. In this case, the quantity $\bar\psi\psi$ can be written as $\bar\psi\psi = \langle\bar\psi\psi\rangle + \sigma$~\cite{Gross:1974jv,WALECKA1974491}, where $\sigma$ is a real scalar field (``fluctuations''), assumed to be small, i.e. $|\sigma / \langle\bar\psi\psi\rangle| \ll 1$, and $\langle\bar\psi\psi\rangle$ is just a constant. Independent of the coordinates, that is. Such a coordinate-independent chiral condensate $\langle\bar\psi\psi\rangle$ is sometimes referred to as {\it global} chiral condensate~\cite{Ohata:2020myj,Carabba:2021xmc}, to be distinguished from the {\it local} chiral condensate $\langle\bar\psi(x)\psi(x)\rangle$ which does depend on the coordinate. The global chiral condensate $\langle\bar\psi\psi\rangle$ will depend on the chemical potential $\mu$ and the temperature $T$. Then, the Lagrangian becomes
\bea
\mathcal L = \bar\psi(i\slashed\partial - m + 2g\langle\bar\psi\psi\rangle + \mu\gamma_0)\psi - g\langle\bar\psi\psi\rangle^2 + \mathcal O(\sigma^2)\,.
\eea
Neglecting $\mathcal O(\sigma^2)$ terms, this is the Lagrangian $\mathcal L_\mathrm{Dirac}$ of a free Dirac fermion (at finite chemical potential) with mass $M = m - 2g\langle\bar\psi\psi\rangle$ with a constant potential $\mathcal V = g\langle\bar\psi\psi\rangle^2 = (M - m)^2/4g$. Thus, we have an effective, linearized, Lagrangian
\bea
\label{Leff}
\mathcal L_\mathrm{eff} = \bar\psi(i\slashed\partial - M + \mu\gamma_0)\psi - \frac{(M - m)^2}{4g} = \mathcal L_\mathrm{Dirac} - \mathcal V\,,
\eea
such that $\mathcal L = \mathcal L_\mathrm{eff} + \mathcal O(\sigma^2)$, and $M = M(\mu, T)$ is the effective mass of the fermion field.

The value of the chiral condensate $\langle\bar\psi\psi\rangle$, or equivalently the effective mass $M$, can then be determined from the condition of thermal equilibrium, i.e. that the Grand Canonical Potential $\Omega(\mu, T; M) = -\frac{T}{L}\ln\mathcal Z$ is minimized\footnote{Here, the volume is the ``volume'' of 1 dimensional space, so it is just a length $L$.}. The partition function $\mathcal Z$ is given, in the path integral formulation, by \cite{Kapusta}
\begin{equation}
    \mathcal Z = \int \mathcal{D}\psi\int \mathcal{D}\bar{\psi}\exp\left[\int_0^\beta d\tau\int dx\, \mathcal L_E \right]\,,
\end{equation}
where the Euclidean Lagrangian $\mathcal L_E$ is obtained from making the change of variables to imaginary time $\tau = it$, and $\beta = 1/T$ is the coldness. In particular, from our effective Lagrangian (\ref{Leff}), we have $\mathcal L_{\mathrm{eff}, E} = \mathcal L_{\mathrm{Dirac}, E} - \mathcal V$ so 
\begin{align}
    \mathcal Z &= \int \mathcal{D}\psi\int \mathcal{D}\bar{\psi}\exp\left[\int_0^\beta d\tau\int dx \,\mathcal L_{\mathrm{eff}, E} \right] \nonumber\\
    &= \int \mathcal{D}\psi\int \mathcal{D}\bar{\psi}\exp\left[\int_0^\beta d\tau\int dx\, (\mathcal L_{\mathrm{Dirac}, E} - \mathcal V) \right] \nonumber\\
    &= e^{-\beta L\mathcal V}\int \mathcal{D}\psi\int \mathcal{D}\bar{\psi}\exp\left[\int_0^\beta d\tau\int dx\, \mathcal L_{\mathrm{Dirac}, E} \right] \nonumber\\
    &= e^{-L\mathcal V/T}\mathcal Z_\mathrm{Dirac} \nonumber\\
    &= e^{-\frac{L}{T}(\Omega_\mathrm{Dirac} + \mathcal V)}\,.
\end{align}
From this, we see that the Grand Canonical Potential $\Omega$ is given by $\Omega = \Omega_\mathrm{Dirac} + \mathcal V$, where $\Omega_\mathrm{Dirac}$ is the Grand Canonical Potential of a free Dirac field of mass $M$ (in one spatial dimension), which is given by \cite{Kapusta,Buballa:2003qv}
\bea
\Omega_\mathrm{Dirac}(\mu, T; M) = -\frac{2}{\pi}\int_0^\infty dk\left[\omega_k + T\ln(1 + e^{-\beta(\omega_k + \mu)})  + T\ln(1 + e^{-\beta(\omega_k - \mu)})\right]\,,
\eea
where $\omega_k = \sqrt{k^2 + M^2}$. Then, by adding the potential $\mathcal V = (M - m)^2/4g$, we obtain
\bea
\label{GC}
\Omega(\mu, T; M) = \frac{(M - m)^2}{4g}-\frac{2}{\pi}\int_0^\infty dk\left[\omega_k + T\ln(1 + e^{-\beta(\omega_k + \mu)})  + T\ln(1 + e^{-\beta(\omega_k - \mu)})\right]\,.
\eea

This quantity diverges, so it must be regularized. However, for the sake of comparison with the Lattice model given in Sec.~\ref{sec:njlqite} below, we have a natural momentum cutoff $\Lambda = \pi/a$, where $a$ is the lattice spacing. By imposing this hard momentum cutoff in the integral of Eq.~\eqref{GC}, we may numerically compute the effective mass $M$ at fixed values of $\mu,~T$ by minimizing $\Omega$ with respect to $M$: $\partial \Omega/\partial M = 0$, which leads to the so-called gap equation. The chiral condensate can then be calculated from $\langle\bar\psi\psi\rangle = (m - M)/2g$. The results obtained this way will be referred to as {\it the analytical calculation} in the following sections. We performed this analysis for $\mu, T$ in the range $[0  \ \mathrm{MeV}, 300 \ \mathrm{MeV}]$ using the model parameters $m = 100$ MeV, $a = 1$ MeV$^{-1}$ and $g=1$ that are relevant for our simulation. The resulting mass surface is plotted in Fig.~\ref{fig:3dplot} below.
\begin{figure}[h]
\centering
\includegraphics[width=0.48\textwidth,trim={0 0.6cm 0 0},clip]{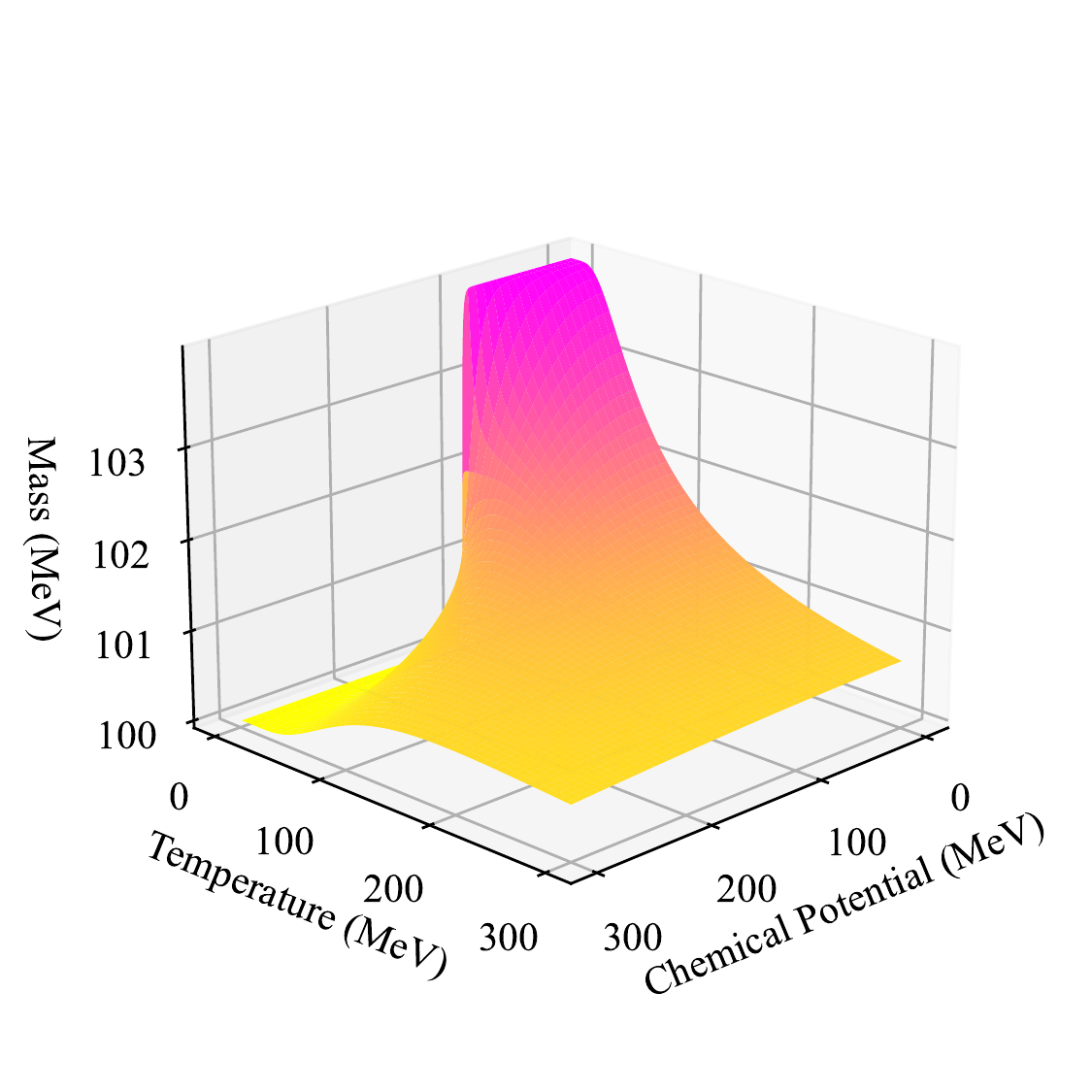}
\includegraphics[width=0.48\textwidth]{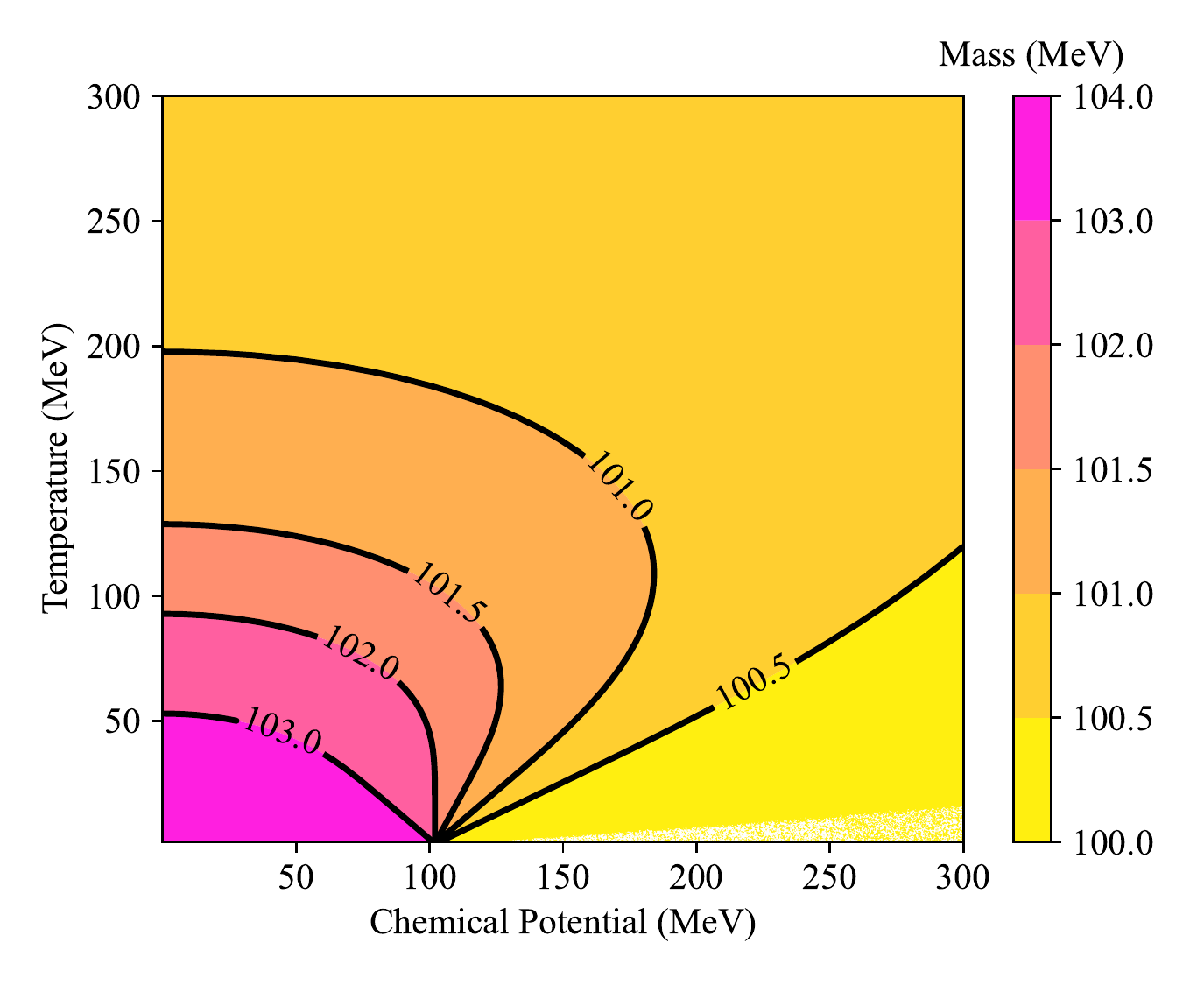}
\caption{The effective mass $M$ as a function of the baryochemical potential $\mu$ and temperature $T$. Left: 3D Surface Plot, Right: Contour Plot.}\label{fig:3dplot}
   \end{figure}
As can be seen, for sufficiently low $\mu^2 + T^2$ there is a dynamically generated mass of around $\Delta m \equiv M-m = 4$ MeV, in addition to the mass $m = 100$ MeV present in the Lagrangian, with $M \rightarrow m$ asymptotically as expected. Historically the GN model was studied specifically because it is \textit{asymptotically free}~\cite{Gross:1974jv}. Thus, at asymptotically high temperatures/chemical potentials we expect to recover a free field theory.
 
\section{Quantum algorithms for the NJL model}\label{sec:njlqite}
In this section, we provide the ingredients of the quantum algorithms for the NJL model, in particular, we first show the discretization of the NJL Hamiltonian on a lattice grid, then we introduce the quantum imaginary time evolution algorithm. 
\subsection{NJL Hamiltonian at the lattice fermion model}\label{sec:njl}
Starting from the Lagrangian density given in Eq.~\eqref{eq:Lagrangian}, one can obtain the NJL Hamiltonian density as follows
 \bea
\label{eq:Hamiltonian}
\mathcal{H}&=\bar{\psi}(i\gamma_1{\partial_1}+m)\psi-g(\bar{\psi}\psi)^2-\mu\bar{\psi}\gamma_0\psi\,.
\eea
Given a Dirac fermion field $\psi(x)$ with components $\rho(x)$ and $\eta(x)$, to discretize the theory we place the fermion field on a spatial lattice, with spacing $a$, and set a staggered fermion field $\chi_n$ on each lattice site~\cite{PhysRevD.16.3031}. Note that staggered fermions are used to avoid a subtlety known as the \textit{fermion doubling problem}, where na\"{i}ve discretization of the fermion field leads to a theory with too many degrees of freedom~\cite{Staggered}. 

With even $N$ and integers $n=0,1,\cdots N/2-1$, sites $2n$ are assigned to the stagger fermion field $\chi_{2n}/\sqrt{a}$ with spinor $\rho(x=n)$. And similarly, sites $2n+1$ are assigned to the stagger fermion field $\chi_{2n+1}/\sqrt{a}$ with spinor $\eta(x=n)$. Then the position $x$ is discretized into $N/2$ position points $x_0,\cdots,x_{N/2-1}$ with spacing $x_{n}-x_{n-1}=\delta x=a$. And the Dirac fermion field $\psi(x)$ at position $x=n$ on the lattice is represented by staggered fermion fields~\cite{Borsanyi:2010cj,Borsanyi:2013bia,Aoki:2005vt,HotQCD:2014kol,Aubin:2019usy},
\bea
\psi(x=n)=\left(\begin{array}{cc}
     &\ \rho(x=n)\  \\
     &\  \eta(x=n)\
\end{array}\right)=\frac{1}{\sqrt{a}}\left(\begin{array}{cc}
     &\ \chi_{2n}\  \\
     &\ \chi_{2n+1}\
\end{array}\right)\,,
\eea 
Also note that the partial derivative of the field $\partial_1\psi(x)$ is defined by the forward-backward derivative convention~\cite{Chakrabarti:2002yu} in lattice QCD and given by
\bea
\partial_1\psi(x=n)=\left(\begin{array}{cc}
     &\ \partial_1\rho(x=n)\  \\
     &\  \partial_1\eta(x=n)\
\end{array}\right)=\frac{1}{a\sqrt{a}}\left(\begin{array}{cc}
     &\ \chi_{2(n+1)}-\chi_{2n}\  \\
     &\ \chi_{2n+1}-\chi_{2(n-1)+1}\
\end{array}\right)\,.
\eea
To show the discretization of the fermion field, we specify $\psi_n$ to represent $\psi(x=n)$, then one can easily find the following equivalence
\bea
&\int dx\,\bar{\psi}\psi=\delta x\sum_{n=0}^{N/2-1}{\psi}_n^\dagger\gamma_0\psi_n=\sum_{n=0}^{N-1}(-1)^n\chi_n^\dagger\chi_n\,,\\
&\int dx\,\bar{\psi}\gamma_0\psi=\delta x\sum_{n=0}^{N/2-1}{\psi}_n^\dagger\psi_n=\sum_{n=0}^{N-1}\chi_n^\dagger\chi_n\,,\\
&\int dx\,\bar{\psi}i\gamma_1\partial_1\psi=\delta x\sum_{n=0}^{N/2-1}{\psi}_n^\dagger i\gamma_5\partial_1\psi_n\nnu
&\hspace{2.3cm}=-\frac{i}{2a}\left\{\left[\sum_{n=0}^{N-2}\left(\chi_n^\dagger\chi_{n+1}-\chi_{n+1}^\dagger\chi_n\right)\right]+\left(\chi_{N-1}^\dagger\chi_{0}-\chi_{0}^\dagger\chi_{N-1}\right)\right\}\,,
\label{eq:bound}
\\
&\int dx\,(\bar{\psi}\psi)^2=\delta x\sum_{n=0}^{N/2-1}({\psi}_n^\dagger\gamma_0\psi_n)^2=\frac{1}{a}\sum_{n=0}^{N/2-1}\left(\chi_{2n}^\dagger\chi_{2n}-\chi_{2n+1}^\dagger\chi_{2n+1}\right)^2\,.
\eea
Keep in mind that $N$ is an even number and thus $N/2$ is always an integer in the above formula. Note that we have taken a periodic boundary condition, namely $\chi_N\to \chi_0$, and the $\chi_{N-1}$ field would be coupled with the $\chi_{0}$ field. For illustration, we have explicitly written this boundary term out in Eq.~\eqref{eq:bound}.
Therefore, the Hamiltonian in spin representation is given by
\bea
H=&\int dx \left[\bar{\psi}(i\gamma_1\partial_1+m)\psi-g(\bar{\psi}\psi)^2-\mu\bar{\psi}\gamma_0\psi]\right]\nnu
=&-\frac{i}{2a}\left[\sum_{n=0}^{N-2}\left(\chi_n^\dagger\chi_{n+1}-\chi_{n+1}^\dagger\chi_n\right)+\left(\chi_{N-1}^\dagger\chi_{0}-\chi_{0}^\dagger\chi_{N-1}\right)\right]+m\sum_{n=0}^{N-1}(-1)^n\chi_n^\dagger\chi_n\nnu
&-\frac{g}{a}\sum_{n=0}^{N/2-1}\left(\chi_{2n}^\dagger\chi_{2n}-\chi_{2n+1}^\dagger\chi_{2n+1}\right)^2-\mu\sum_{n=0}^{N-1}\chi_n^\dagger\chi_n\,.\label{Hamiltonia_all}
\eea
Once one converts the Dirac fermions to staggered fermions, by applying the Jordan-Wigner transformation \cite{Jordan1928}, 
\bea
\chi_n=\frac{\sigma_{X,n}-i\sigma_{Y,n}}{2}\prod_{i=0}^{n-1}\left(-i\sigma_{Z,i}\right)\,,\label{eq:jw-trans}
\eea
one obtains the operators in spin representations for quantum simulation. In Eq.~\eqref{eq:jw-trans}, $\sigma_{X,n}$ represents a Pauli-$X$ matrix acting on the $n$-th grid on the lattice, etc. 
Then one is able to verify the following equivalence up to a constant term, 
\bea
&\int dx\,\bar{\psi}i\gamma_1\partial_1\psi=\sum_{n=0}^{N-2}\frac{1}{4a}\left(\sigma_{X,{n}}\sigma_{X,{n+1}}+\sigma_{Y,{n}}\sigma_{Y,{n+1}}\right)\nnu
&\hspace{2.8cm}+\frac{(-1)^{N/2}}{4a}\left(\sigma_{X,{N-1}}\sigma_{X,{0}}+\sigma_{Y,{N-1}}\sigma_{Y,{0}}\right)\prod_{i=1}^{N-2}\sigma_{Z,i}\,,
\label{eq:wboundary}
\\
&\int dx\,\bar{\psi}\psi=\sum_{n=0}^{N-1}(-1)^n\frac{\sigma_{Z,n}}{2}\,,
\label{eq:barpsipsi}
\\
&\int dx\,\bar{\psi}\gamma_0\psi=\sum_{n=0}^{N-1}\frac{\sigma_{Z,n}}{2}\,,
\\
&\int dx\,(\bar{\psi}\psi)^2=-\frac{1}{4a}\Bigg\{\bigg[\sum_{n=0}^{N-2}(\mathbb{1}+\sigma_{Z,n})(\mathbb{1}+\sigma_{Z,n+1})\bigg]+(\mathbb{1}+\sigma_{Z,N-1})(\mathbb{1}+\sigma_{Z,0})\Bigg\}
\nonumber\\
&\hspace{24mm}+\frac{1}{2a}\sum_{n=0}^{N-1}\left(\mathbb{1}+\sigma_{Z,n}\right)\,,
\label{eq:bound2}
\eea
where the subscripts indicate the index of qubit where the single-qubit gate is acting on. Here again we have explicitly written out the boundary term in Eqs.~\eqref{eq:wboundary} and \eqref{eq:bound2}. Note that even though Eq.~\eqref{eq:bound2} has a slightly different form, it is equivalent to the expression given in~\cite{PhysRevD.16.3031}. In other words, by applying the Jordan-Wigner transformation from Eq.~\eqref{eq:jw-trans} on Eq.~\eqref{Hamiltonia_all}, we obtain the Hamiltonian $H=\sum_{j=1}^5 H_j$ decomposed into the following 5 pieces for constructing the quantum algorithm,
\bea
H_1=&\,\frac{m}{2}\sum_{n=0}^{N-1}(-1)^n\sigma_{Z,n}\,,
\label{Hamiltonia3}
\\
H_2=&\,-\frac{\mu}{2}\sum_{n=0}^{N-1}\sigma_{Z,n}\,,
\label{Hamiltonia5}
\\
H_3=&\sum_{n=0}^{N/2-1}\frac{1}{4a}\left(\sigma_{X,{2n}}\sigma_{X,{2n+1}}+\sigma_{Y,{2n}}\sigma_{Y,{2n+1}}\right)\,,
\label{Hamiltonia1}
\\
H_4=&\sum_{n=1}^{N/2-1}\frac{1}{4a}\left(\sigma_{X,{2n-1}}\sigma_{X,{2n}}+\sigma_{Y,{2n-1}}\sigma_{Y,{2n}}\right)\nnu
&\,+\frac{(-1)^{N/2}}{4a}\left(\sigma_{X,{N-1}}\sigma_{X,{0}}+\sigma_{Y.{N-1}}\sigma_{Y,{0}}\right)\prod_{i=1}^{N-2}\sigma_{Z,i}\,,
\label{Hamiltonia2}
\\
H_5=&\frac{g}{4a}\bigg[\sum_{n=0}^{N-2}(\mathbb{1}+\sigma_{Z,n})(\mathbb{1}+\sigma_{Z,n+1})+(\mathbb{1}+\sigma_{Z,N-1})(\mathbb{1}+\sigma_{Z,0})-2\sum_{n=0}^{N-1}(\mathbb{1}+\sigma_{Z,n})\bigg]\,,
\label{Hamiltonia4}
\eea
With these Hamiltonians, one is able to apply Suzuki-Trotter decomposition~\cite{trotter1959product,Suzuki} and evolve the Hamiltonian using a quantum simulation.

\subsection{Quantum imaginary time evolution algorithm}\label{sec:qite}
To perform a quantum simulation for calculating the thermal properties of the NJL Hamiltonian in Eq.~\eqref{eq:Hamiltonian} at finite temperatures, we use the quantum imaginary time evolution (QITE) algorithm developed in \cite{Motta}. Quantum simulation of a Hamiltonian \(H\) traditionally uses the Suzuki-Trotter decomposition to approximate the evolution \(e^{-iHt}\ket{\Psi}\) by a series of local unitary operations which are implementable on a quantum computer. In QITE, the imaginary time substitution \(\beta = it\) is first made, followed by a series of decomposition and approximation steps.

QITE is especially useful as a subroutine within the quantum minimally entangled typical thermal states (QMETTS) algorithm \cite{Motta}, which is used to calculate thermal averages of observables. 
The benefit of using imaginary time evolution to calculate thermal averages is that, compared to other quantum algorithms \cite{2011,chowdhury16,brandao19} for computing thermal averages, QITE does not require deep circuits or any ancilla qubits.
We will use QITE and QMETTS to calculate the chiral condensate $\langle\bar\psi\psi\rangle$ for the purpose of studying chiral phase transition.

The QITE algorithm aims to decompose the imaginary time evolution operator \(e^{-\beta H}\). Taking the Hamiltonian as given in Sec.~\ref{sec:njl}, Trotterization of the imaginary time evolution operator yields
\bea
e^{- \beta H} = \left(e^{-\Delta\beta H}\right)^{n} + O(\Delta\beta^2),
\eea
where \(\Delta\beta\) is the chosen imaginary time step size and \(n = \beta / \Delta\beta\) is the number of iterations needed to reach imaginary time \(\beta\).

\begin{figure}[h]
\centering
\includegraphics[width=0.78\textwidth]{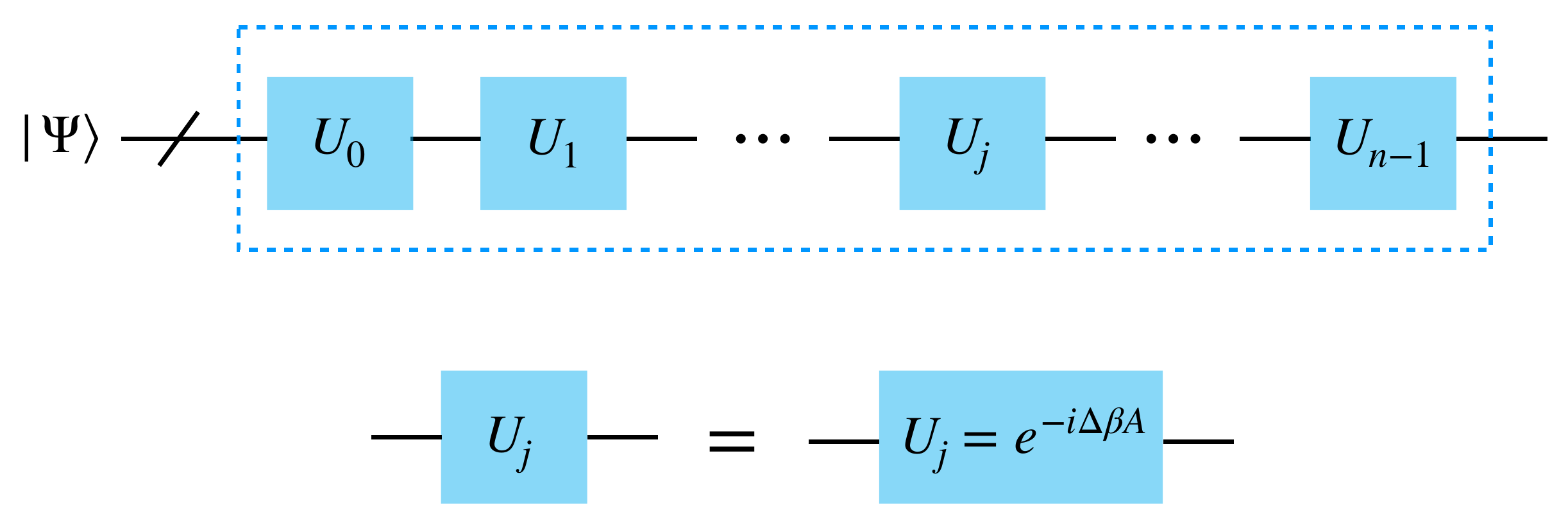}
\caption{Quantum circuit for generating the thermal state at $\beta=n\Delta\beta$ evolved from $\ket{\Psi}$. Each block evolves the state with a small imaginary time interval $\Delta\beta$ and all $n$ blocks inside the dashed-line box form the QITE algorithm for thermal state preparation.}\label{fig:init}
\end{figure}

In contrast with real-time quantum simulation, each intermediate evolution \(e^{- \Delta\beta H}\) is non-unitary, and cannot be directly implemented in terms of quantum gates. The crux of QITE lies in approximating the non-unitary evolution with a unitary operation so that (with proper normalization $c(\Delta\beta)=\langle\Psi|e^{-2\Delta\beta H}\ket{\Psi}$)
\bea
\frac{1}{\sqrt{c(\Delta\beta)}} e^{- \Delta\beta H} \ket{\Psi} \approx  e^{-i \Delta\beta A} \ket{\Psi},\label{eq:nonu_to_u}
\eea
where $A$ is a Hermitian operator parameterized with a linear combination of Pauli operators,
\bea
A(\bm a)=\sum_{\mu}a_\mu\hat{\sigma}_{\mu}\,, {\quad {\rm with}~} \hat{\sigma}_{\mu} = \prod_{l}{\sigma}_{\mu_l}^{(l)}\,,\label{eq:q6}
\eea
where $a_\mu$ are coefficients to be determined below and $\hat{\sigma}_{\mu}$ is short-hand notation for the Pauli string given above. The subscript $\mu$ in $a_\mu$ and $\hat{\sigma}_{\mu}$ is a composite index $\mu=(\mu_1,\mu_2,\cdots,\mu_{n_q})$ with $\mu_l\in\{I,\,X,\,Y,\,Z\}$ and thus runs over all possible $4^{n_q}$ subsets of Pauli strings. On the other hand, $\sigma_{\mu_l}^{(l)}$ represents a Pauli-$\mu_l$ operator acting on the $l$-th qubit in the circuit. In this work, we choose the number of qubits $n_q=4$. When $\Delta\beta$ is very small, one can expand both sides of Eq.~\eqref{eq:nonu_to_u} up to $\mathcal{O}(\Delta\beta)$
\bea
\left[\mathbb{1}+ \left(\frac{1-\sqrt{c(\Delta\beta)}-\Delta\beta H}{\sqrt{c(\Delta\beta)}}\right)\right] \ket{\Psi} \approx  [\mathbb{1}-i \Delta\beta A] \ket{\Psi}\,.
\eea
Thus the change rates of the quantum states under $e^{-\beta H}$ and $e^{-i\beta A}$ at imaginary time $\beta$ are given by
\bea
\ket{\Delta\Psi_H(\beta)}&=\big({\ket{\Psi(\beta+\Delta\beta)}-\ket{\Psi(\beta)}}\big)/{\Delta\beta}\nnu
&=\left(\frac{1}{\sqrt{c(\Delta\beta)}} e^{- \Delta\beta H}-\mathbb{1}\right)\ket{\Psi(\beta)}/{\Delta\beta}\nnu
&\approx\left(\frac{1-\sqrt{c(\Delta\beta)}-\Delta\beta H}{\Delta\beta\sqrt{c(\Delta\beta)}}\right)\ket{\Psi(\beta)}\,,\\
\ket{\Delta\Psi_A(\beta)}&=\big({\ket{\Psi(\beta+\Delta\beta)}-\ket{\Psi(\beta)}}\big)/{\Delta\beta}\nnu
&=\big( e^{- i\Delta\beta A}-\mathbb{1}\big)\ket{\Psi(\beta)}/{\Delta\beta}\nnu
&\approx -iA\ket{\Psi(\beta)}\,.
\eea
To evaluate the Hermitian operator $A$, we need to minimize the objective function $F(a)$ to obtain the values of $a_{\mu}$. Here $F(a)$ is given by
\bea
F(a)=&||\big(\ket{\Delta\Psi_H(\beta)}-\ket{\Delta_A\Psi(\beta)}\big)||^2\nnu
=&\bra{\Delta\Psi_H(\beta)}\Delta\Psi_H(\beta)\rangle+i\bra{\Delta\Psi_H(\beta)}A\ket{\Psi(\beta)}-i\bra{\Psi(\beta)}A^\dagger\ket{\Delta\Psi_H(\beta)}\nnu
&+\bra{\Psi(\beta)}A^\dagger A\ket{\Psi(\beta)}\nnu
=&F(0)+i\sum_\mu \frac{a_\mu}{\sqrt{c(\Delta\beta)}}\bra{\Psi(\beta)} \big(H\hat{\sigma}_\mu-\hat{\sigma}_\mu^\dagger H\big)\ket{\Psi(\beta)}+\sum_{\mu,\nu}a_\nu a_\mu\bra{\Psi(\beta)}\hat{\sigma}_\nu^\dagger\hat{\sigma}_\mu\ket{\Psi(\beta)}\,,
\label{eq:fa}
\eea
where $F(0)=||\,\ket{\Delta\Psi_H(\beta)}\,||^2$ corresponds to $F(a)$ at $a=0$, thus this term is independent of the values of $a_\mu$. To minimize the objective function $F(a)$, one is required to take a derivative of $a$ and solve the linear equation below
\bea
&\frac{i}{\sqrt{c(\Delta\beta)}}\bra{\Psi(\beta)} \big(H\hat{\sigma}_\mu-\hat{\sigma}_\mu^\dagger H\big)\ket{\Psi(\beta)}+\sum_\nu a_\nu\bra{\Psi(\beta)}\big(\hat{\sigma}_\nu^\dagger\hat{\sigma}_\mu+\hat{\sigma}_\mu^\dagger\hat{\sigma}_\nu\big)\ket{\Psi(\beta)}=0\,.\label{eq:linear0}
\eea
Therefore, by defining the elements of matrix ${\bm S}$ and vector ${\bm b}$ as
\bea
S_{\mu\nu}&=\bra{\Psi(\beta)}\hat{\sigma}_\mu^\dagger\hat{\sigma}_\nu\ket{\Psi(\beta)}\,,
\\
b_\mu&=-\frac{i}{\sqrt{c(\Delta\beta)}}\bra{\Psi(\beta)} \big(H\hat{\sigma}_\mu-\hat{\sigma}_\mu^\dagger H\big)\ket{\Psi(\beta)}\,,
\eea
one obtains the parameters $a_\mu$ by solving the linear equation 
\bea
({\bm S}+{\bm S}^T){\bm a}={\bm b}\,.
\eea
 With $a_\mu$ obtained, the unitary evolution \(e^{-i \Delta\beta A}\) is implementable in terms of quantum gates. We illustrate this implementation in Fig.~\ref{fig:init}, where $U_j|\Psi(\beta_j)\rangle= e^{-i\Delta\beta A}|\Psi(\beta_j)\rangle=|\Psi(\beta_j+\Delta\beta)\rangle$. Namely, each block evolves state $\ket{\Psi(\beta_j)}$ to $\ket{\Psi(\beta_j+\Delta\beta)}$, and after $n$ steps, one obtains the state at the target total evolution time $\beta$. 

Using the thermal state $|\Psi(\beta/2)\rangle$ generated by the QITE algorithm, 
we then find the expectation value of an observable $\hat{O}$ at finite temperature $T=1/\beta$, which can be written as
\bea
\langle \hat{O}\rangle_\beta=\frac{\langle\Psi(\beta/2)|\hat{O}|\Psi(\beta/2)\rangle}{||\ |\Psi(\beta/2)\rangle||^2}=\frac{{\rm Tr}(e^{-\beta \hat{H}}\hat{O})}{{\rm Tr}(e^{-\beta \hat{H}})}\,.\label{eq:operator}
\eea
Here the two traces in the numerator and denominator are computed as follows
\bea
{\rm Tr}(e^{-\beta \hat{H}}\hat{O})&=\sum_{i\in\mathcal{S}}\bra{i}e^{-\beta \hat{H}/2}\hat{O}e^{-\beta \hat{H}/2}\ket{i}\,,
\label{eq:TrO}
\\
{\rm Tr}(e^{-\beta \hat{H}})&=\sum_{i\in\mathcal{S}}\bra{i}e^{-\beta \hat{H}}\ket{i}\,,
\eea
where $\mathcal{S}$ is the complete set of $2^{n_q}$-dimensional Hilbert space. In other words, $\mathcal{S} = \{|0000\rangle,\ |0001\rangle,\ |0010\rangle,\cdots,|1111\rangle\}$ with 16 orthogonal basis for $n_q=4$ qubits used in our quantum simulation below. Below, we will compute the expectation value of the {\it global} chiral condensate $\langle\bar\psi\psi\rangle$. In doing so, we choose the operator $\hat O$ in Eq.~\eqref{eq:TrO} with $\hat O =\int dx\, \bar{\psi}(x)\psi(x)/\int dx$ and applying Eq.~\eqref{eq:barpsipsi}. In other words, we take the average of $\bar{\psi}(x)\psi(x)$ on all the lattice sites.

\begin{figure}[h]
\centering
\includegraphics[width=\textwidth]{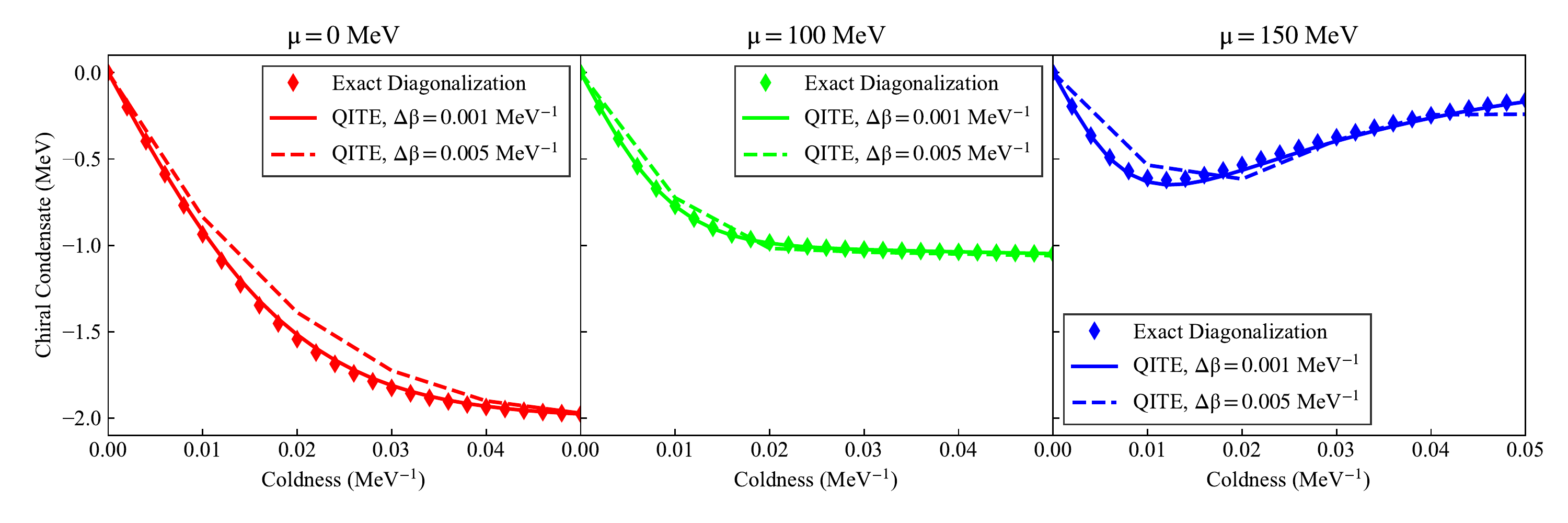}
\caption{Chiral condensate $\langle\bar{\psi}\psi\rangle$ at chemical potentials $\mu=0$ MeV (left), $100$ MeV (middle) and $\mu=150$ MeV (right) versus coldness $\beta=1/T$ (MeV$^{-1}$) simulated for $(1+1)$-dimensional NJL model with coupling constant $g=1$, bare quark mass $m=100$ MeV and lattice spacing $a=1$ MeV$^{-1}$. We present the simulation results with imaginary time step $\Delta\beta=0.001$ in solid lines and $\Delta\beta=0.005$ in dashed lines. For comparison, the exact diagonalization results are shown in points.}\label{fig:sec2mu}
\end{figure}

In Fig.~\ref{fig:sec2mu}, we present the imaginary time evolution of the chiral condensate $\langle\bar\psi\psi\rangle$ at chemical potentials $\mu=0$ MeV (left), $100$ MeV (middle) and $\mu=150$ MeV (right) with small-time step $\Delta\beta=0.001$ MeV$^{-1}$ (solid line) and $\Delta\beta=0.005$ MeV$^{-1}$ (dashed line). Our quantum algorithm contains 4 qubits. The quantum simulation is performed on a $(1+1)$-dimensional NJL model with the dimensionless coupling constant $g=1$ and the quark mass $m=100$ MeV on a lattice grid with spacing $a=1$ MeV$^{-1}$. The exact diagonalization carried out with discretization of the NJL Hamiltonian as given in Eq.~\eqref{Hamiltonia_all} is shown in diamond points for reference. As expected, one obtains more accurate quantum simulations with smaller evolution steps. In the following sections, we will apply small-time step $\Delta\beta=0.001$ MeV$^{-1}$ for the QITE algorithm and compare the effective mass $M$ among quantum simulation, exact diagonalization, and theoretical analysis.

\section{Results}\label{sec:pheno}

In this section, we demonstrate the $(1+1)$-dimensional NJL chiral phase transition by plotting the quark condensate $\langle\bar{\psi}\psi\rangle$ as a function of temperature $T$ and chemical potential $\mu$. By working on a $(1+1)$-dimensional model, from the theoretical point of view, we are able to derive an analytical result by solving the gap equation as illustrated in Sec.~\ref{sec:theory} as a comparison to the quantum simulations. At the same time, restriction on $(1+1)$ dimension enables us to design a quantum circuit with a small amount of qubits, which leads to feasibility of implementing such circuit on currently available hardwares for further research.

The parameters that will be used in our work are set as: lattice spacing $a=1$ MeV$^{-1}$, bare mass $m=100$ MeV and the dimensionless coupling constant $g=1$. Our results are obtained from the following three directions: 
\begin{enumerate}
    \item Simulation of the thermal states with the QITE algorithm;
    \item Exact diagonalization of the Hamiltonian in spin representation;
    \item Analytical calculation by numerically solving the gap equation.
\end{enumerate}

To obtain the quantum simulation results, one can install a quantum simulation package such as PYQUILL~\cite{https://doi.org/10.48550/arxiv.1608.03355}, TEQUILA~\cite{Kottmann_2021}, QFORTE~\cite{stair2021qforte}, QISKIT~\cite{gadi_aleksandrowicz_2019_2562111}, XACC~\cite{McCaskey_2020}, CIRQ~\cite{cirq_developers_2021_5182845}, and so on to apply the quantum circuits. These quantum simulation packages give similar results for quantum simulations. They are python software libraries and are used for providing expected outputs of an ideal quantum computer. Algorithm availability for some packages is discussed in~\cite{Anand:2021xbq} and a more complete list of general quantum simulation packages can be found in~\cite{Bharti:2021zez}. In this work, we work within the framework of the open-source software package QFORTE~\cite{stair2021qforte}, which provides an efficient state simulator and quantum algorithms library. We modify the implementation of QITE within QFORTE to calculate properties of the NJL Hamiltonian in quantum simulation. 

\begin{figure}[h]
\centering
\includegraphics[width=0.75\textwidth]{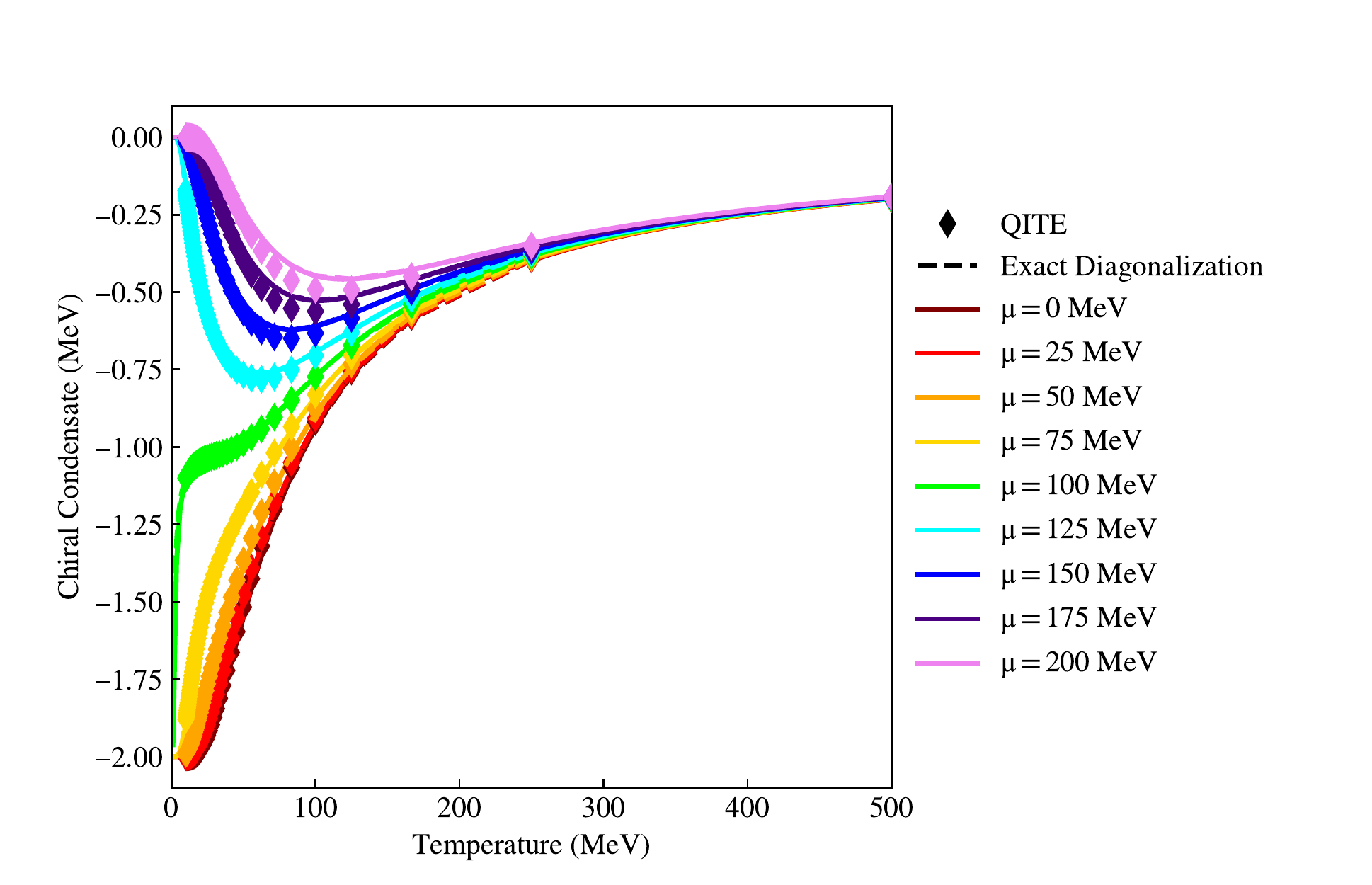}
\caption{Chiral Condensate $\langle
\bar\psi\psi\rangle$ as a function of temperature $T$ at chemical potentials $\mu=0, ..., 200$ MeV. Dashed curves are from exact diagonalization and solid curves are calculated analytically. }\label{fig:pheno1}
\end{figure}

In Fig.~\ref{fig:pheno1}, we plot the temperature dependence of the chiral condensate $\langle\bar{\psi}\psi\rangle$ at chemical potentials set at $\mu=0,\ 25,\cdots,\ 200$ MeV. The quantum simulation results are presented in comparison with theoretical calculations and exact diagonalization. The diamond data points are from QITE simulation, while the solid curves are from the theory calculation as explained in Sec.~\ref{sec:theory}. On the other hand, the dashed curves are from the exact diagonalization which is performed by applying the matrix form of the discretized NJL Hamiltonian given in Eqs.~\eqref{Hamiltonia3} - \eqref{Hamiltonia4}. We find that our quantum simulations are in good agreement with both exact diagonalization and theoretical analyses. For chemical potentials below 100 MeV, the quark condensate increases as the temperature rises. While for large chemical potentials, the quark condensate shows opposite behavior at a small temperature range then slowly increases with temperature as indicated by the figure. And the quark condensate at various finite chemical potentials $\mu$ tends to converge at large temperatures.

\begin{figure}[h]
\centering
\includegraphics[width=0.75\textwidth]{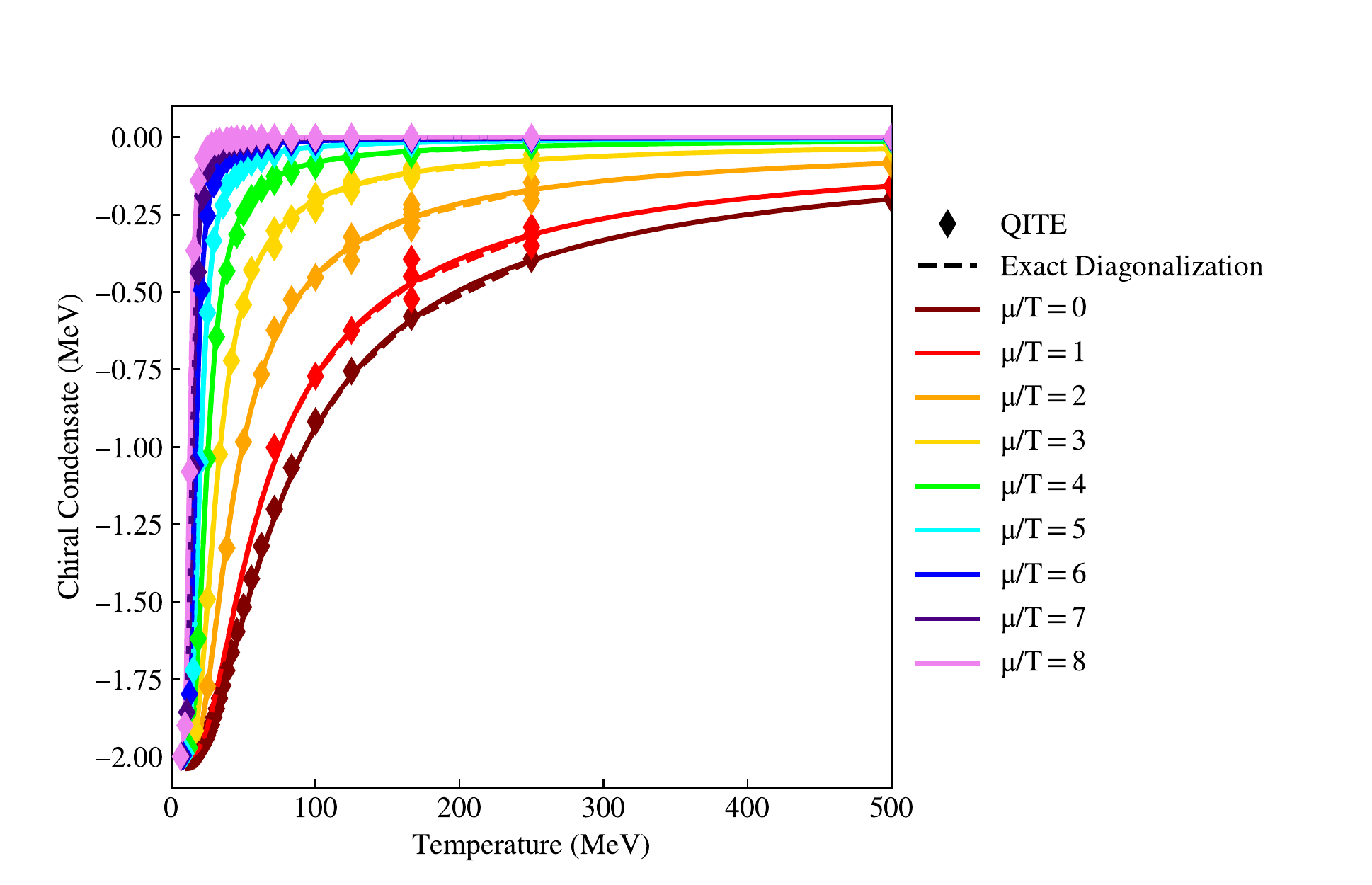}
\caption{Chiral Condensate $\langle
\bar\psi\psi\rangle$ as a function of temperature $T$ at ratios $\mu/T=0, ..., 8$. Dashed curves are from exact diagonalization and solid curves are calculated analytically. }\label{fig:pheno2}
\end{figure}

In Fig.~\ref{fig:pheno2}, we plot the quark condensate $\langle\bar{\psi}\psi\rangle$ as a function of temperature for different ratios of the chemical potential and temperature, $\mu/T$, from 0 to 8 with the same set of parameters chosen for Fig.~\ref{fig:pheno1}. Here the quantum simulations using the QITE algorithm are plotted in diamond points. For comparison, we also provide the analytical calculation and exact diagonalization results as shown in solid lines and dashed lines respectively. At low temperatures, the quark condensate is about $-2$ MeV. As the temperature increases, the quark condensate increases to around $0$ MeV. As indicated by the curves, at a large ratio $\mu/T$, the quark condensate increases from $-2$ to 0 MeV more rapidly.

\begin{figure}[h]
\centering
\includegraphics[width=0.75\textwidth]{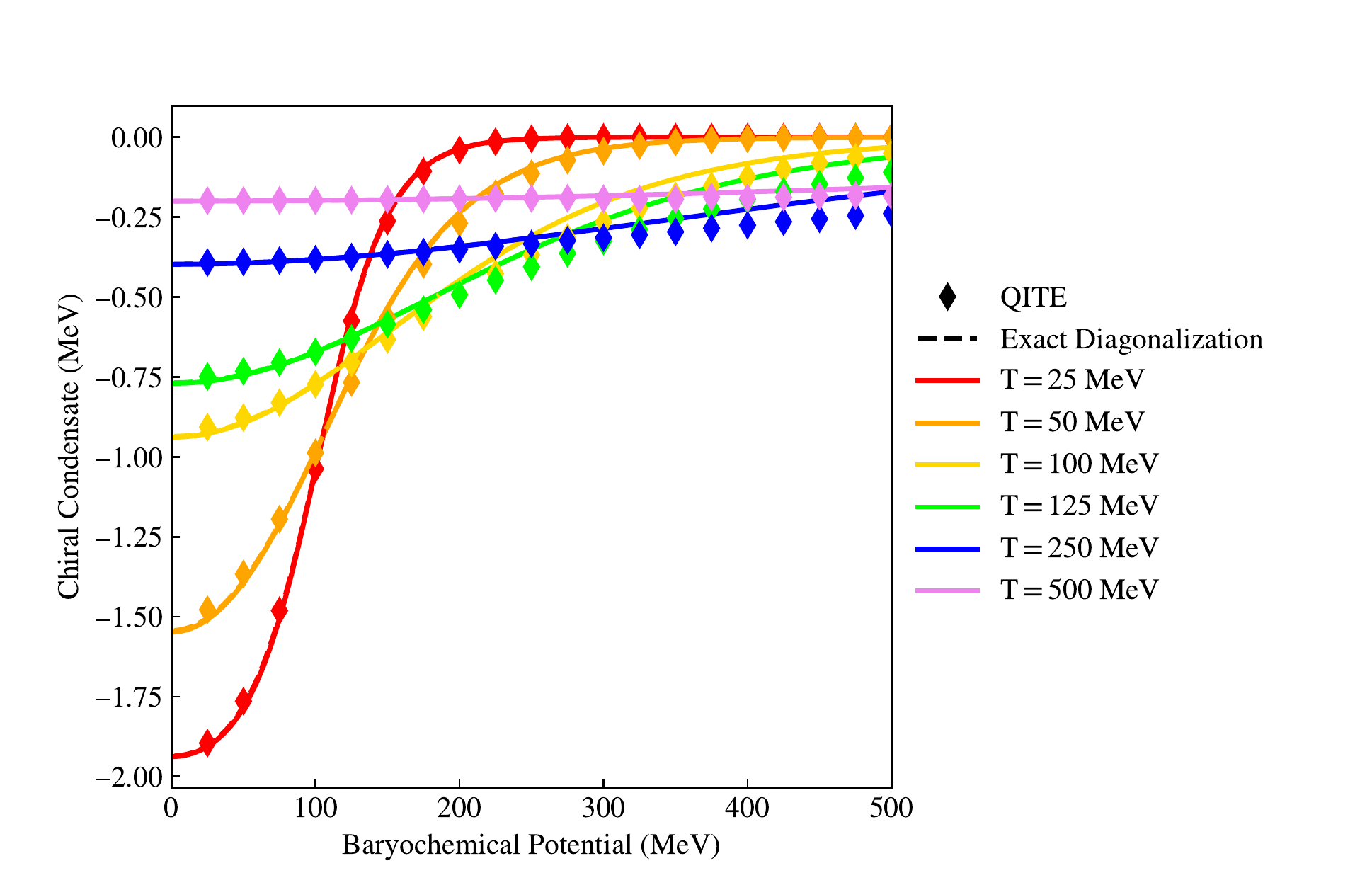}
\caption{Chiral Condensate $\langle
\bar\psi\psi\rangle$ as a function of chemical potential $\mu$ at temperatures $T=25,\ 50,\ 100, 125, 250$ and $500$ MeV. Dashed curves are from exact diagonalization and solid curves are calculated analytically.}\label{fig:pheno3}
\end{figure}

Similarly, we plot the quark condensate $\langle\bar{\psi}\psi\rangle$ for finite chemical potentials at various temperatures in Fig.~\ref{fig:pheno3}, for $T=25,\ 50,\ 100,\ 125,\ 250$ and $500$ MeV in the $\langle\bar{\psi}\psi\rangle-\mu$ plane. At low temperatures and chemical potentials, the quark condensate has a larger absolute value. As the chemical potential $\mu$ increases, the quark condensate rapidly increases to around 0 MeV. While for high temperatures, the quark condensate barely changes at different chemical potentials. Therefore, one observes a more obvious phase transition at larger temperatures in the $\langle\bar{\psi}\psi\rangle-\mu$ plot. 

\section{Conclusion}\label{sec:conclusion}
In this work, we have constructed a quantum simulation for the chiral phase transition of the 1+1 dimensional NJL model at finite temperature and chemical potentials with the QITE algorithm, where we applied a 4-qubit quantum circuit, and simulated the NJL Hamiltonian with quantum gates. We observe a consistency among digital quantum simulation, exact diagonalization and analytical solution, which indicates rich applications of quantum computing in simulating finite-temperature behaviors for QCD in the future. 

The demonstrated insensitivity of the quantum simulation's efficacy to the value of the chemical potential (and likely other external parameters) presents an exciting avenue for exploring finite density effects in QCD and other field theories. This aspect of quark physics remains largely unexplored (in comparison to other areas) due to technical constraints (nonperturbative dynamics, the failure of Monte Carlo due to the sign problem, etc.) preventing the use of traditional computational methods (perturbation theory, Lattice Gauge Theory, Semi-Classical Methods, etc.). With scalable quantum computer technology on the horizon, performing Lattice QCD calculations on quantum computers is not only increasingly possible, but practical as well.

This work, as well as previous studies, demonstrates that NISQ quantum computers are capable of producing consistent and correct answers to physical problems, some of which cannot be efficiently or effectively solved with classical computing algorithms. This signals bright prospects for future applications to nonperturbative QCD and beyond.

\acknowledgments
This work is supported by the National Science Foundation under grant No.~PHY-1945471.

\bibliographystyle{JHEP}
\bibliography{njlqc}

\end{document}